\documentclass[aps,prl,reprint,superscriptaddress,nobibnotes]{revtex4-2}

\usepackage{graphicx}
\usepackage{dcolumn}
\usepackage{bm}
\usepackage{xcolor}
\usepackage{subfig}
\usepackage[justification=centerlast,format=plain]{caption}
\usepackage{tikz}
\usetikzlibrary{positioning,shapes}
\usepackage{footmisc}
\usepackage{threeparttable}
\usepackage[normalem]{ulem}

\newcommand{\LSUAffiliation}{Dept. of Physics and Astronomy, Louisiana State University, Baton Rouge, LA 70803, USA}

\newcommand{\TAMU}
{Cyclotron Institute, Texas A\&M University, College Station, Texas 77843, USA}

\newcommand{\FSU}
{Physics Department, Florida State University, Tallahassee,
  FL 32306, USA}

\newcommand{\WoodsHole}
{Woods Hole Oceanographic Institute, Woods Hole, MA 02543, USA}

\newcommand{\Argonne}
{Physics Division, Argonne National Laboratory, Lemont, IL 60439, USA}

\newcommand{\TRIUMF}
{TRIUMF, 4004 Wesbrook Mall, Vancouver, British Columbia, V6T 2A3, Canada}

\newcommand{\SCElectric}
{S\&C Electric Company, Chicago, IL 60626, USA}

\newcommand{\LANL}
{Los Alamos National Laboratory, P.O. Box 1663, Los Alamos, NM 87545, USA}

\newcommand{\UTK}
{Department of Physics and Astronomy, University of Tennessee, Knoxville, TN 37996, USA}

\newcommand{\SmithsDetection}
{Smiths Detection, 94405 Vitry-sur-Seine, France}

\newcommand{\NRIA}
{National Research and Innovation Agency (BRIN), Jakarta Pusat 10340, Indonesia}

\newcommand{\Helion}
{Helion Energy, Everett, WA 98203, USA}

\newcommand{\Axcelis}
{Axcelis Technologies Inc., Beverly, MA 01915, USA}

\newcommand{\Archbold}
{Archbold Medical Center, Thomasville, GA 31792, USA}

\newcommand{\sizeA}{0.95}

\begin{document}

\preprint{APS/123-QED}

\title{Properties of states in $^{19}$Ne important for the $^{18}$F($p,\alpha$)$^{15}$O reaction rate}

\author{K. H. Pham}
\email{khang\_pham@tamu.edu}
\affiliation{\LSUAffiliation}
\affiliation{\TAMU}

\author{D. Mumma}
\affiliation{\LSUAffiliation}

\author{C. M. Deibel}
\affiliation{\LSUAffiliation}

\author{L. T. Baby}
\affiliation{\FSU}

\author{J. C. Blackmon}
\affiliation{\LSUAffiliation}

\author{K. D. Launey}
\affiliation{\LSUAffiliation}

\author{K. T. Macon}
\affiliation{\LSUAffiliation}
\affiliation{\WoodsHole}

\author{G. W. McCann}
\affiliation{\FSU}
\affiliation{\SCElectric}

\author{B. Sudarsan}
\affiliation{\LSUAffiliation}
\affiliation{\FSU}

\author{I. Wiedenh\"over}
\affiliation{\FSU}

\author{S. Ajayi}
\affiliation{\FSU}
\affiliation{\Archbold}

\author{C. Benetti}
\affiliation{\FSU}
\affiliation{\Helion}

\author{A. Bhardwaj}
\affiliation{\LSUAffiliation}

\author{W. Braverman}
\affiliation{\LSUAffiliation}

\author{K. S. Davis}
\affiliation{\LSUAffiliation}
\affiliation{\TRIUMF}

\author{J. C. Esparza}
\affiliation{\FSU}

\author{K. Hanselman}
\affiliation{\FSU}
\affiliation{\LANL}

\author{D. He}
\affiliation{\LSUAffiliation}

\author{S. Lopez-Caceres}
\affiliation{\LSUAffiliation}
\affiliation{\Axcelis}

\author{E. Lopez-Saavedra}
\affiliation{\FSU}
\affiliation{\Argonne}

\author{M. McLain}
\affiliation{\LSUAffiliation}

\author{A. B. Morelock}
\affiliation{\FSU}
\affiliation{\UTK}

\author{V. Sitaraman}
\affiliation{\FSU}

\author{E. Temanson}
\affiliation{\FSU}
\affiliation{\SmithsDetection}

\author{C. Wibisono}
\affiliation{\FSU}
\affiliation{\NRIA}

\date{\today}

\begin{abstract}
Observation of the 511-keV positron-annihilation line would be a powerful probe of classical novae, with the primary source of positrons likely from the $\beta^+$ decay of \textsuperscript{18}F. 
We have determined the properties of important resonances in $^{19}$Ne, which govern the 
\textsuperscript{18}F($p,\alpha$)\textsuperscript{15}O reaction rate and the production of \textsuperscript{18}F in novae.
Measured $\alpha$ and proton angular distributions from states populated in the \textsuperscript{19}F(\textsuperscript{3}He,$t$)\textsuperscript{19}Ne
reaction identified six near-threshold proton $s$-wave \textsuperscript{18}F$+p$ ($L_p=0$) states, and the asymptotic normalization of these states was studied using the symmetry-adapted no-core shell model. Combining these results with existing direct reaction measurements, we place additional constraints on the uncertainties in the $^{18}$F($p,\alpha$)$^{15}$O reaction rate. Our findings improve the current understanding of the states contributing to this rate and demonstrate that previous studies significantly underestimated its uncertainties.
\end{abstract}

\maketitle



Classical novae are thermonuclear explosions that occur in close binary stellar systems where hydrogen-rich material accretes onto a white dwarf. These astrophysical phenomena are the primary Galactic generators of some stable isotopes (\textsuperscript{13}C, \textsuperscript{15}N, and \textsuperscript{17}O) and can produce many long-lived radioactive isotopes, e.g., \textsuperscript{18}F, \textsuperscript{22}Na, and \textsuperscript{26}Al, which emit $\gamma$ rays \cite{Jose2006}. Analysis of nova light curves and $\gamma$-ray signatures may constrain the properties of the nova, the binary system, and the accompanying nucleosynthesis. At this time, only $>$100-MeV $\gamma$ rays have been observed well after the nova eruption, but current models predict stronger emissions occurring within hours after eruption \cite{Fermi-LAT2014, Cheung2016, Hernanz1999, Jose2006}. This early emission is expected to be dominated by the 511-keV line from positron annihilation following $\beta^+$ decay of \textsuperscript{18}F ($t_{1/2}=110$ min) and, to a lesser extent, \textsuperscript{13}N ($t_{1/2}=862$ s). Early detectability of novae therefore hinges on production of \textsuperscript{18}F. The predicted early 511-keV emission has eluded detection to date, making improved nuclear constraints on \textsuperscript{18}F destruction essential to interpret forthcoming MeV transient searches. One aim of the upcoming Compton Spectrometer and Imager (COSI) mission is to measure the luminosity of $\gamma$ rays from classical novae, and specifically the 511-keV line \cite{COSI}. This presents an exciting opportunity to constrain 
these astrophysical phenomena and their contribution to Galactic chemical evolution.

The amount of \textsuperscript{18}F in the envelope of a nova is particularly sensitive to its rate of destruction by the \textsuperscript{18}F($p,\alpha$)\textsuperscript{15}O reaction \cite{Coc2000}. This reaction occurs via resonances in the compound nucleus \textsuperscript{19}Ne, and resonance energies, strengths, and widths of excited states in \textsuperscript{19}Ne are needed to determine the \textsuperscript{18}F($p,\alpha$)\textsuperscript{15}O reaction rate. A recent analysis of the properties of important states in $^{19}$Ne is provided in Ref. \cite{Kim25}. Previous examinations of the \textsuperscript{18}F($p,\alpha$)\textsuperscript{15}O astrophysical $S$-factor and reaction rate have also shown that significant uncertainty arises from
interference effects between 
$J^\pi=\frac{1}{2}^+,\frac{3}{2}^+$ states near the proton threshold in \textsuperscript{19}Ne ($S_p=6410$ keV) that have an $L=0$ ($s$ wave) \textsuperscript{18}F$+p$ entrance channel.

 A sub-threshold proton $s$-wave state at 6.286 MeV was identified in the $^{20}$Ne($p,d$)$^{19}$Ne reaction, and interference effects among the 6.286-, 6.457-, 7.075-, and 7.871-MeV states were studied in Ref. \cite{Bardayan2015} using information from previous studies (e.g., Ref. \cite{Adekola2011}) resulting in a significantly reduced uncertainty in the reaction rate. More recently,  Kahl \emph{et al.} identified additional proton $s$-wave levels, including sub-threshold states and proton-unbound resonances (5.345, 5.486, 5.824, 6.130, 6.421, 7.088, and 7.79 MeV) and studied interference effects of the 6.130- and 6.421-MeV states near the proton threshold \cite{Kahl2019, Kahl2021}. The addition of these proton $s$-wave levels, along with variations in the signs of the reduced partial widths and the proton and alpha widths, resulted in a greater uncertainty in the $S$-factor, especially at lower energies. Subsequently, Portillo \emph{et al}. confirmed the proton $s$-wave assignments for the 6.13- and 7.79-MeV states, and while the uncertainties in the $S$-factor near the proton threshold were within previously reported values, those at energies well above the proton threshold were substantially greater than previous determinations \cite{Portillo2023}. These results underscore that an accurate reaction rate requires accurate identification of proton $s$-wave  levels and their interference.


We measured the \textsuperscript{19}F(\textsuperscript{3}He,$t$)\textsuperscript{19}Ne transfer reaction to study states in \textsuperscript{19}Ne at the John D. Fox Superconducting Linear Accelerator Laboratory at Florida State University.
A 24-MeV $^3$He beam with intensities between $12-40$ enA bombarded a 120-$\mu$g/cm\textsuperscript{2} CaF\textsubscript{2} target on a 20-$\mu$g/cm\textsuperscript{2} carbon backing. Tritons were momentum analyzed using the Super Enge Split-Pole Spectrograph (SE-SPS) positioned at an angle of 3\textdegree\ with respect to the beam axis. This forward-angle geometry was chosen to strongly suppress nonzero magnetic-substate contributions, which vanish at 0\textdegree, thereby simplifying the calculation of the decay angular correlations \cite{Visser2004}. Horizontal and vertical slits were open to an effective SE-SPS entrance aperture of 4.62 msr. The magnetic field of the spectrograph was set to 12 kG for detection of excited states in \textsuperscript{19}Ne with $E_x>4.0$ MeV. Tritons were detected by a position-sensitive proportional counter backed by a scintillator at the focal plane of the SE-SPS. Triton events were isolated through various particle identification plots including residual energy $(E_r)$ vs. energy loss $(\Delta E)$, position $(x)$ vs. $E_r$, and $x$ vs. $\Delta E$. Further position-corrections and energy calibrations were applied to ($^{3}$He,$t$) events.

\begin{figure}[!b]
        \centering
	\includegraphics[width=\sizeA\linewidth]{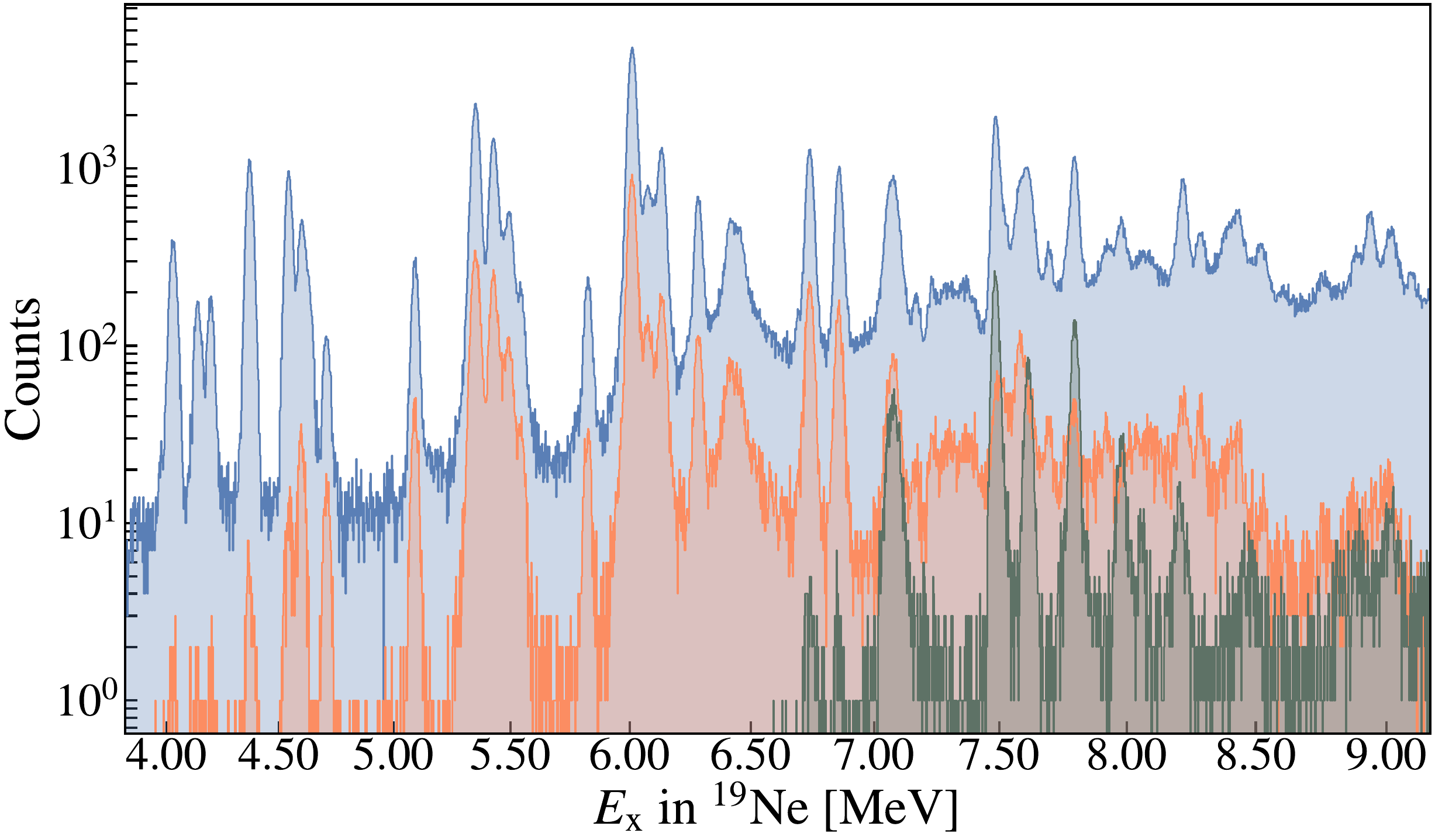}
	\caption{
		Calibrated spectra of all triton events at the focal plane (blue) and those triton events in coincidence with proton (black) and $\alpha$ (red) decays detected in SABRE.
		\label{fig:coincident}
	}
\end{figure}

 The Silicon Array for Branching Ratio Experiments (SABRE) was placed at backward angles in the target chamber to detect light ions from the decays of unbound states populated in \textsuperscript{19}Ne in coincidence with tritons at the focal plane. Signals in SABRE were calibrated using channel-by-channel and global gain matching.
Coincident events between SABRE and the focal plane were identified by synchronizing all SABRE channels in time with the scintillator signals. Events within a 140-ns coincidence timing window were selected, and background contributions were estimated and subtracted using timing windows placed just outside the coincidence region. The excitation energy spectra for total triton singles and coincident events are shown in Fig. \ref{fig:coincident}. For a more detailed description of both the focal plane detector and SABRE calibrations, and the data processing, see Ref. \cite{Phamthesis, Good2021}.

The quantities of observed tritons corresponding to each state in $^{19}$Ne were determined by analyzing the focal plane spectrum, with each peak modeled as a Gaussian on a linear background. The signals from SABRE were divided into angular bins according to the center-of mass emission angle $\theta_{\text{c.m.}}$ of the detected decay particle relative to the beam axis, each combining counts from consecutive detector strips. 
Histograms were generated for triton events in coincidence with each angular bin, determining the number of decay counts per solid angle $dn/d\Omega$ for individual states in $^{19}$Ne after correcting for geometric efficiency. The angular dependence of the decay was then characterized by fitting the extracted $dn/d\Omega$ values with a Legendre polynomial expansion of the form:
\begin{equation}
    W(\theta)=\sum_{k = 0,2,4,...}^{2L} A_k P_k(\cos\theta) .
\end{equation}
\newline
\noindent
The quality of the fit $(\chi^2)$ and $A_0$ coefficients were used to determine $L$-values, and angular distributions were integrated to determine branching ratios 
(${\Gamma_x}/{\Gamma}$) 
for each decay channel, $x$. 
Angular distributions of select states are shown in Fig. \ref{fig:angular_distributions}.  

In this work, we have identified six proton $s$-wave levels between approximately 5 and 7 MeV: five sub-threshold (proton-bound) states at 5.548, 5.826, 6.067, 6.133, and 6.285 MeV, and one proton-unbound resonance at 7.074 MeV. Results for these and other states of particular interest for the \textsuperscript{18}F($p,\alpha$)\textsuperscript{15}O reaction are summarized in Table \ref{results}. For the proton-bound states, only the $E_x$ = 6.285-MeV state has an experimentally constrained proton asymptotic normalization coefficient ($p$-ANC), for which we adopt the value from Ref. \cite{Adekola2011}. For the remaining proton-bound states, $p$-ANCs were calculated theoretically in this work using the ab initio symmetry-adapted no-core shell model with continuum (SA-NCSM) \cite{Launey2021, LAUNEY2026104233}. Proton-\textsuperscript{18}F overlap functions, $\langle$\textsuperscript{19}Ne$|p$+\textsuperscript{18}F(g.s.)$\rangle$, were computed using the NNLOopt chiral potential in 9 major shells, and the corresponding $s$-wave $p$-ANCs were calculated following the matching procedure described in Ref. \cite{Sargsyan2023}. Uncertainties are reported for variations in the $\hbar \Omega$ basis parameter (from 11 to 15 MeV) with a 10\% contribution stemming from the underlying nuclear interaction, as suggested in Ref. \cite{BeckerLED22}. As a cross-check, the SA-NCSM calculation gave a $p$-ANC of 74(8) fm\textsuperscript{-1/2} for the 6.285-MeV state, consistent with the experimentally deduced value of 83 fm\textsuperscript{-1/2} from Ref. \cite{Adekola2011}, which we adopt in the $R$-matrix analysis. Below we discuss states that significantly impact the reaction rate and/or have discrepant assignments previously reported. 

\renewcommand{\arraystretch}{1.75}

\begin{table}[!b]
  \caption{Properties of states in \textsuperscript{19}Ne. Values are from this work except where noted. Signs and uncertainties (in parenthesis) for the proton ANC (fm\textsuperscript{-1/2}) for proton-bound states or $\Gamma_{p}$ (keV) for unbound states are given in the fifth column and were varied in the $R$-matrix analysis. For levels included in the $R$-matrix calculations, the listed 2$J^\pi$ is the value adopted in the analysis. Where two 2$J^\pi$ assignments are shown, the present data do not discriminate between them and those levels were not included in the $R$-matrix model.}
  \label{results}

\begin{threeparttable}

    \begin{ruledtabular}
      \begin{tabular}{cccccc}
        $E_x$ (MeV) & $L_\alpha$ & 2$J^\pi$ & $\Gamma_\alpha/\Gamma$ & $\Gamma_p/\Gamma$ & $\Gamma_p$; ANC \\
        \colrule

        4.031(3) & & $3^+$\tnote{a} & 1\tnote{a} & & $(\pm)1.1(6)$ \\
        5.352(3) & 1 & $1^+$ & 1\tnote{b} & & $(\pm)7(1)$ \\
        \ldots & & & & \\

        5.491(3) & 2 & $3^{\boldsymbol{-}},5^{\boldsymbol{-}}$ & 0.71(4) & & \\
        5.548(3) & 1 & $3^+$ & 0.72(11) & & $(\pm)3(2)$ \\
        5.826(3) & 1 & $1^+$ & 1.10(8) & & $(\pm)4(1)$ \\
        \ldots & & & & & \\

        6.067(3) & 1 & $3^+$ & 0.79(7) & & $(\pm)3(2)$ \\
        \ldots & & & & & \\

        6.133(3) & 1 & $3^+$ & 0.86(5) & & $(\pm)7(4)$ \\
        6.285(3) & 1 & $1^+$ & 0.88(8) & & $(\pm)83\tnote{c}$ \\
        6.416(2) & 2 & $3^{\boldsymbol{-}}$ & & & $5\times 10^{-50}\tnote{f}$\\
        6.436(2) & 3 & $5^+,7^+$ & & & \\
        6.440(2) & 2 & $3^{\boldsymbol{-}},5^{\boldsymbol{-}}$ & & & \\
        6.459(2) & 1 & $3^+$ & & & $(\pm 2\times 10^{-13})\tnote{d}$ \\
        6.740(2) & 2 & $3^{\boldsymbol{-}}$ & 0.90(12) & & $2.2(7)\times 10^{-3}\tnote{e}$ \\
        \ldots & & & & & \\

        7.074(2) & 1 & $3^+$ & 0.69(4) & 0.36(2) & $(\pm)15\tnote{d}$ \\
        \ldots & & & & & \\

        7.615(2) & 1 & $3^+$ & 0.41(4) & 0.45(2) & $(\pm)14$ \\
        7.796(2) & 1 & $1^+$ & 0.40(2) & 0.66(6) & $(\pm)16$ \\
        7.835\tnote{a} & & $1^+$\tnote{a} & & & $(\pm)60\tnote{a}$ \\

      \end{tabular}
    \end{ruledtabular}

    {
      \footnotesize
      \begin{tabular}{lll}
        \textsuperscript{a} Ref.~\cite{Kim25}. & \textsuperscript{c} Ref.~\cite{Adekola2011}. & \textsuperscript{e} Ref.~\cite{Bardayan2002}. \\
        \textsuperscript{b} Used for normalization. & \textsuperscript{d} Ref.~\cite{Bardayan2015}. & \textsuperscript{f} Ref.~\cite{Hall2019}. \\
      \end{tabular}
    }

  \end{threeparttable}
\end{table}


\begin{figure*}[t]
    \centering
    \includegraphics[width=\sizeA\linewidth]{lValuesSelectedGrid_paper.png}
    \caption{
        Angular distribution fits for select excited states in \textsuperscript{19}Ne. $\theta_{\text{c.m.}}$ is the center-of-mass emission angle of the detected decay particle ($\alpha$) relative to the beam axis. For each angular bin, the plotted $\theta_{\text{c.m.}}$ is taken as the arithmetic mean of the center-of-mass angles of the detector strips included in that bin.
        \label{fig:angular_distributions}
    }
\end{figure*}

Kahl \emph{et al.} identified a candidate sub-threshold proton $s$-wave state at 5.486 MeV \cite{Kahl2019}; however, our alpha angular distribution analysis indicates $L_\alpha=2$ ($J^\pi=\frac{3}{2}^{\boldsymbol{-}},\frac{5}{2}^{\boldsymbol{-}}$) for a resonance at 5.491 MeV, while we instead identify a state at 5.548 MeV as being proton $s$ wave ($J^\pi=\frac{1}{2}^+,\frac{3}{2}^+$) (Fig. \ref{fig:angular_distributions}).

The assignment of state(s) near 6.285 MeV has been questionable, with studies providing evidence for a doublet. Parikh \emph{et al.} \cite{Parikh2015} fit this region with two unresolved contributions based on the relatively broad width observed.
Strong $\gamma$-ray branching observed to the 4.634-MeV ($J^\pi=\frac{13}{2}^+$) state in another measurement indicates the presence of a state with high spin, likely $J^\pi=\frac{11}{2}^+$ \cite{Hall2019}.
Kahl \emph{et al.} were not able to observe a pure Gamow-Teller transition of this state, suggesting that their peak was an unresolved doublet of both a proton $s$-wave level and a non-$s$-wave level \cite{Kahl2019}. Portillo \emph{et al.} were also unable to fit their angular distribution data to a single $L$ value, and instead used a mixture of $J^\pi=\frac{7}{2}^+$ and $\frac{1}{2}^+$ \cite{Portillo2023}.

As shown in Fig. \ref{fig:angular_distributions}, our alpha angular distribution analysis favors a single $L_\alpha=1$ $(J^\pi=\frac{1}{2}^+,\frac{3}{2}^+)$ fit for the 6.285-MeV excited state, which is consistent with previous results found by Adekola \emph{et al}. and Bardayan \emph{et al}. \cite{Adekola2011, Bardayan2015}. From the angular distribution data and measured width of 33-keV FWHM, we do not see evidence of a doublet, but we cannot rule out this possibility given that Parikh \emph{et al.} proposed the width of this structure (whether a doublet or singlet) as $\approx$16 keV \cite{Parikh2015}. However, if a higher-spin state exists as suggested by Ref. \cite{Portillo2023, Kahl2019}, we find it to be weakly populated, and only the $J^\pi=\frac{1}{2}^+,\frac{3}{2}^+$ state will impact the \textsuperscript{18}F($p,\alpha$)\textsuperscript{15}O reaction rate.

The $E_x = 6.4 - 6.5 \text{-MeV}$ region consists of both narrow, unresolved resonances and a broad resonance.
We find a broad state at 6.436 MeV requiring $L_\alpha \geq 3$ $(J^\pi=\frac{5}{2}^+,\frac{7}{2}^+$ or perhaps $\frac{7}{2}^-,\frac{9}{2}^-)$. This state was previously assigned $\frac{5}{2}^+$  as the mirror to the 6.54-MeV state in $^{19}$F determined from  $\alpha$-scattering measurements \cite{Gol22,Cog19}, but it will not contribute significantly to the \textsuperscript{18}F$(p,\alpha)$\textsuperscript{15}O reaction rate with any of these $J^{\pi}$ assignments.

\begin{figure}[!t]
    \includegraphics[width=\sizeA\linewidth]{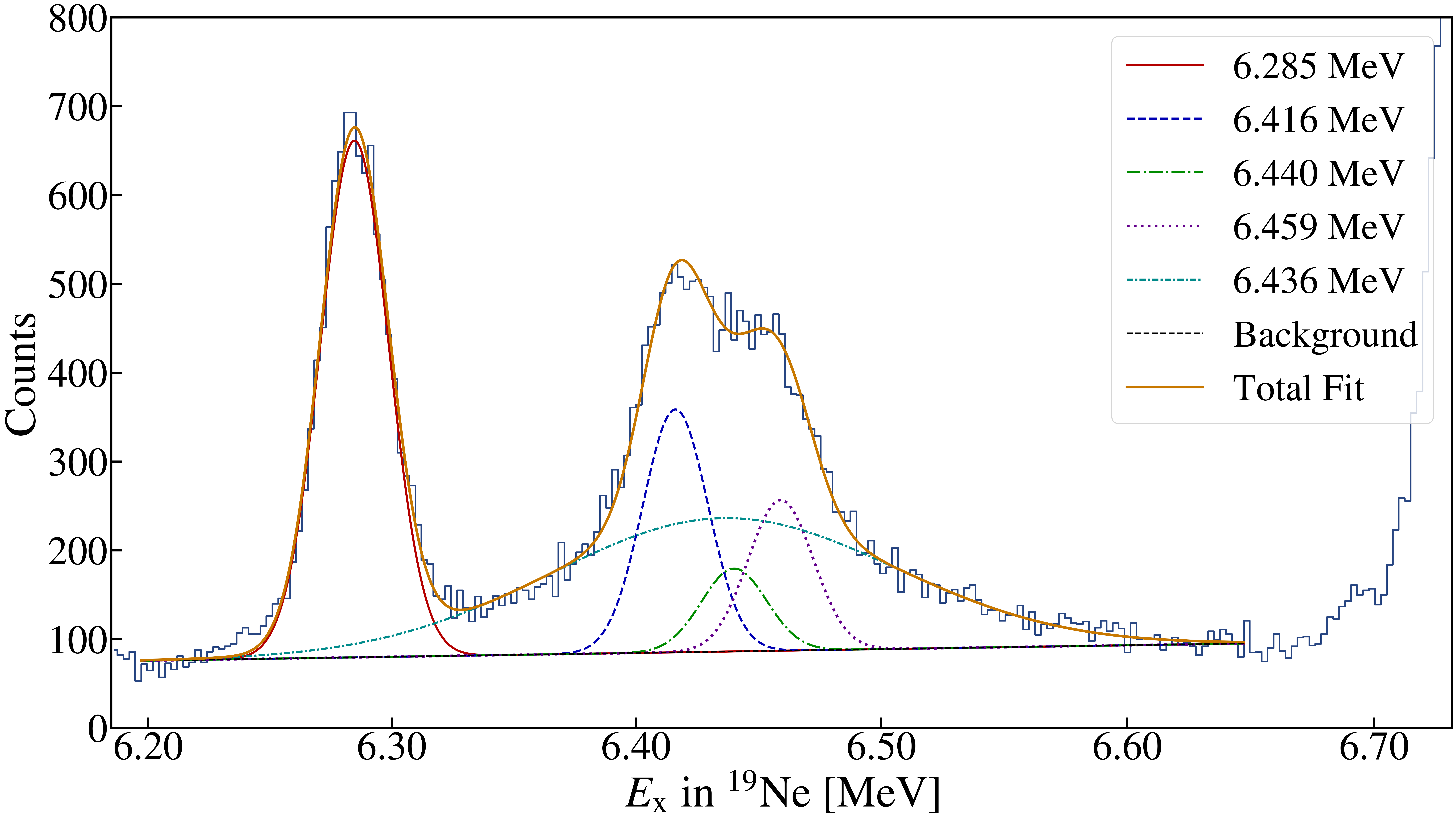} \\
    \caption{
	Constrained fit to the excitation function using fixed widths and centroids following Ref. \cite{Laird2013}.
	\label{fig:fitting_6.4_6.5}
    }
\end{figure}

While the narrow resonances in this region of our data can be well fit using only two states, 
other high-resolution studies utilizing the \textsuperscript{19}F(\textsuperscript{3}He,$t$)\textsuperscript{19}Ne reaction provide evidence for three narrow resonances in this region \cite{Utku1998, Laird2013, Parikh2015}. We therefore adopted a fitting procedure similar to that of Laird \emph{et al.} \cite{Laird2013}, in which we fixed the widths of three peaks in this region to be equal to that of narrow and isolated peaks in our spectra (i.e., our experimental resolution of 31-keV FWHM) and fixed the centroids to be equal to the excitation energies measured by Ref. \cite{Laird2013}. In addition to 
three narrow peaks, the broad peak was also included in the fit without constraints on the width or energy (Fig. \ref{fig:fitting_6.4_6.5}). 

Alpha angular distribution fits favor $L_\alpha=2$ for all three narrow resonances, corresponding to $J^\pi=\frac{3}{2}^{\boldsymbol{-}} \text{ or } \frac{5}{2}^{\boldsymbol{-}}$. While $L_\alpha = 2$ provides the best fit for the 6.459-MeV excited state, the other two narrow resonances are also clearly $L_\alpha = 2$, and their widths overlap, as shown in Fig.~\ref{fig:fitting_6.4_6.5}. Consequently, the $L_\alpha = 2$ assignment for the 6.459-MeV state may reflect contamination from the lower-energy states included in the fit, and an $L_\alpha = 1$ assignment for the 6.459-MeV state (corresponding to $L_p=0$) is not statistically ruled out by our data (Fig. \ref{fig:angular_distributions}).  

Resonances well above the proton threshold may also influence the reaction rate due to potential interference effects with lower-lying resonances. Two such resonances have been identified in this work at $E_x$ = 7.615 and 7.795 MeV, which lie in a sparsely-studied energy region. The angular distributions of the proton decays for the 7.615-MeV excited state are best fit by $L_p=0$ $(J^\pi=\frac{1}{2}^+,\frac{3}{2}^+)$ \cite{Phamthesis}. The 7.795-MeV excited-state proton decays are best-fit with $L_p=2$ $(J^\pi=\frac{1}{2}^+,\frac{3}{2}^+,\frac{5}{2}^+,\frac{7}{2}^+)$, while the alpha decays are best fit with $L_\alpha=1$ $(J^\pi=\frac{1}{2}^+,\frac{3}{2}^+)$ \cite{Phamthesis}. Therefore, we also adopt $J^\pi=\frac{1}{2}^+,\frac{3}{2}^+$ for this state.

\begin{figure}
        \centering
	\includegraphics[width=\sizeA\linewidth]{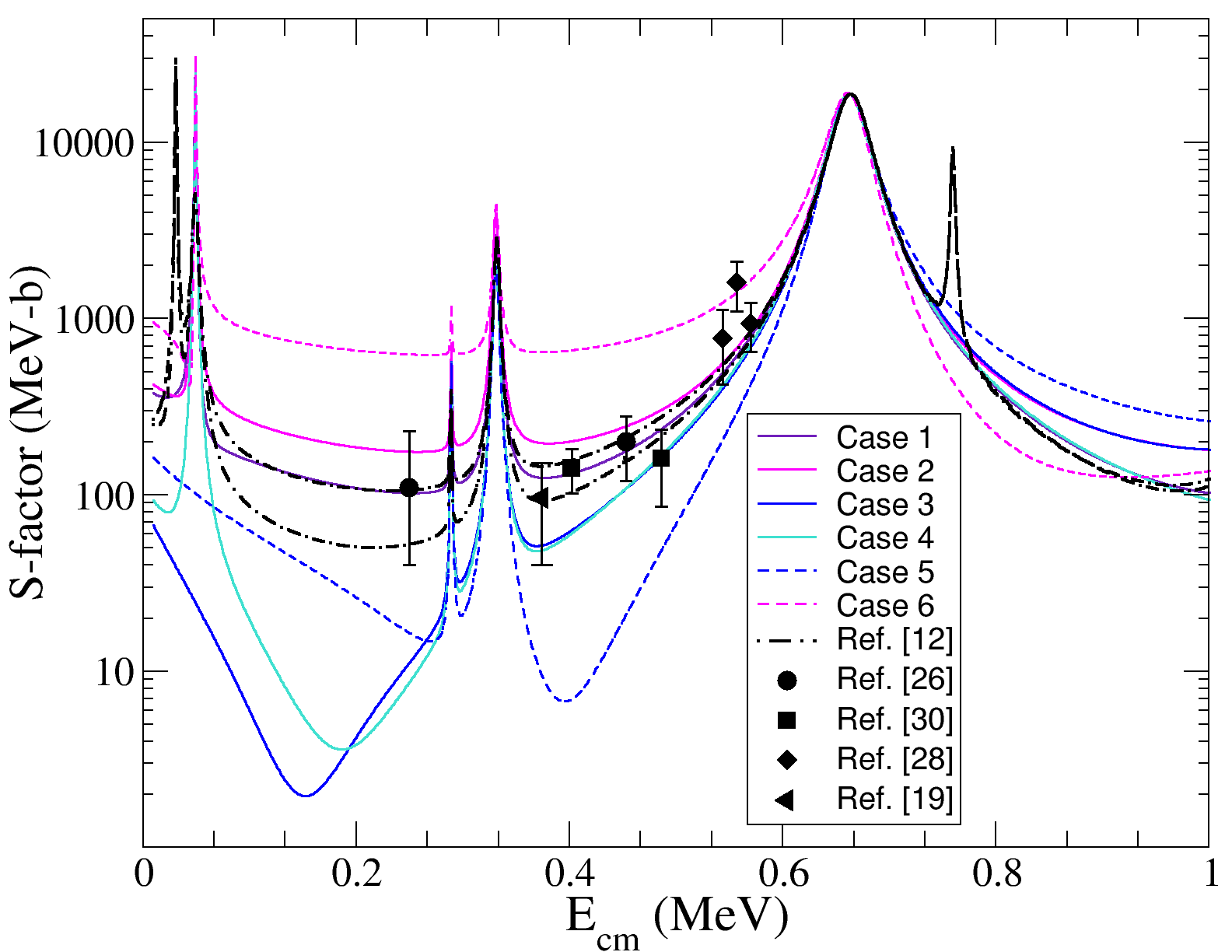}
	\caption{
		Sample astrophysical $S$-factor calculations in AZURE2 for the \textsuperscript{18}F$(p,\alpha)$\textsuperscript{15}O reaction based on our results (colored lines) are compared to those from Ref. \cite{Portillo2023} (black dashed lines) and to direct cross section data from Ref. \cite{Beer2011,Sereville2009, Bardayan2001, Bardayan2002}. For a full description of Cases 1 - 6, see the Supplemental Material.}
		\label{fig:sfactors}
	
\end{figure}

\begin{figure}[!t]
        \centering
	\includegraphics[width=\sizeA\linewidth]{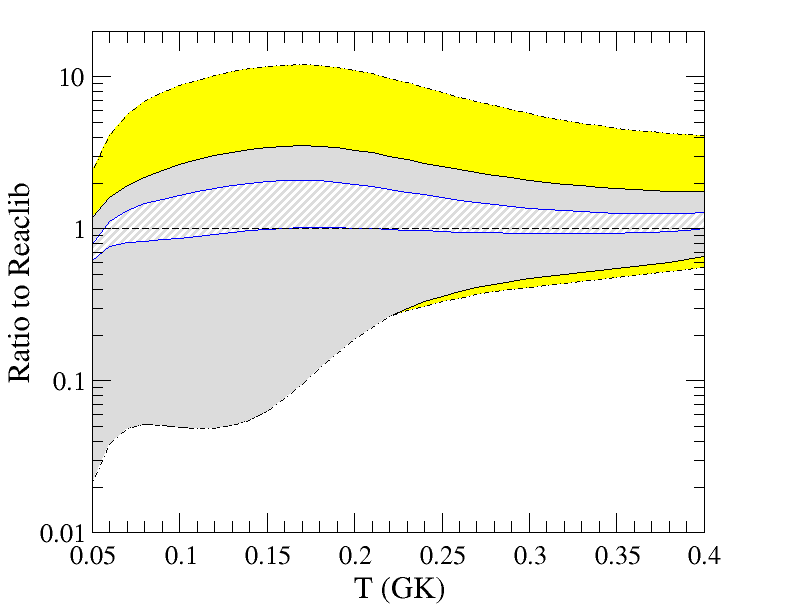}
	\caption{
		Ratios of \textsuperscript{18}F$(p,\alpha)$\textsuperscript{15}O reaction rates to Reaclib \cite{REACLIB}.  The upper and lower bounds shown as black dashed lines (and within the yellow shading) include all rates from this work. The solid black lines (and gray shading) indicate the range of rates consistent with direct \textsuperscript{18}F$(p,\alpha)$\textsuperscript{15}O cross section measurements as described in the text. The blue lines (and hatched region) show the range of rates from Ref. \cite{Portillo2023} for comparison.}
		\label{fig:rates}
	
\end{figure}

To evaluate the impact of 
the identified \textsuperscript{18}F+$p$ $s$-wave levels, we calculated the \textsuperscript{18}F$(p,\alpha)$\textsuperscript{15}O astrophysical $S$-factor and reaction rate using AZURE2 \cite{Azuma2010}. Proton ANCs for bound proton $s$-wave states were varied within their uncertainties shown in Table I. 
Calculations were performed both with and without a 6.459-MeV state with $L_p = 0$ and resonance parameters consistent with that of Ref. \cite{Bardayan2015}.
Signs of reduced partial widths were varied for bound as well as resonant states. We also varied the proton width for the 6.740-MeV state within its uncertainty as it contributes significantly to the $S$-factor in the region of interest despite not being a proton $s$-wave state. Values for parameters for all higher energy states ($E_x>6.750$~MeV) were held fixed in our analysis. While the properties of these states introduce additional uncertainty, we want to clearly illustrate the uncertainty arising only from the states we studied. Six example cases are shown in Fig. \ref{fig:sfactors} that illustrate that the $(p,\alpha)$ cross section at energies important for novae (0.1 - 0.4 GK) is highly sensitive to uncertainties in ANCs for the bound proton $s$-wave states, which have not previously been studied rigorously. The range of reaction rates that result from the cases considered is shown in Fig. \ref{fig:rates}. Much greater uncertainty arises from interference than was previously estimated from studies that considered interference from only a limited subset of states near the proton threshold, e.g. Refs. \cite{Bardayan2015, Portillo2023}. Notably, these greater uncertainties do not result from greater experimental uncertainty, but rather from the interference among the near-threshold proton $s$-wave states identified in this work, together with the remaining uncertainties in their $p$-ANCs and relative reduced-width signs.

While only the 6.285-MeV state has direct experimental constraints on its ANC, constraints on interference between states come from previous off-resonance \textsuperscript{18}F($p,\alpha$)\textsuperscript{15}O cross section measurements in the $E_\mathrm{cm}=250-700$ keV region above the proton threshold \cite{Bardayan2001, Bardayan2002, DeSereville2009, Beer2011}. As shown in Fig. \ref{fig:sfactors}, some cases are ruled out by this existing data. To estimate the uncertainty in the reaction rate consistent with current data, we considered all calculations that have a total chi-squared of $\chi^2<\chi_{\rm min}^2+8$, considering the 8 data points shown. More precise data from direct measurements of the \textsuperscript{18}F($p,\alpha$)\textsuperscript{15}O reaction at higher energies were not included in the current analysis as that energy region is influenced by many higher-lying states with uncertain properties.  Cases 1-4, shown in Fig. \ref{fig:sfactors}, all fall within this range, while Cases 5 and 6 (shown as dashed lines) are examples of cases excluded by this criterion. The direct \textsuperscript{18}F$(p,\alpha)$\textsuperscript{15}O cross section data greatly reduces the upper range of allowed reaction rates as shown in Fig. \ref{fig:rates}. However, the lower range of rates is much less constrained due to the large uncertainty in smaller cross sections (note the logarithmic $y$-axis in Fig. \ref{fig:sfactors}). The uncertainty in reaction rates from a recent analysis using a more limited set of states is shown in Fig. \ref{fig:rates} for comparison \cite{Portillo2023}.



These results show that the near-threshold $s$-wave level structure constrained by the present measurements permits a substantially broader range of interferences than has been considered in more restricted rate determinations. The resulting spread in uncertainties reflects those in the $p$-ANCs and relative reduced-width signs, while existing \textsuperscript{18}F($p,\alpha$)\textsuperscript{15}O cross-section data exclude some of the allowed interference patterns. Specifically, the results of this work imply that the $^{18}$F($p,\alpha$)$^{15}$O reaction rate could be significantly lower than the most recent determination by Portillo \emph{et al.}, resulting in an increased abundance of \textsuperscript{18}F and, consequently, a higher predicted flux of 511-keV gamma rays. This outcome would make the detection of fainter novae via gamma-ray telescopes possible. As is illustrated by the results in Figs. \ref{fig:sfactors} and \ref{fig:rates}, accurate knowledge of the spin-parity assignments and the proton ANCs for the sub- and near-threshold region is critical, as is the interference between multiple states that is likely only constrained by better cross section measurements in the non-resonant region.

\begin{acknowledgments}
 This work was partially supported by the U.S. Department of Energy, Office of Science under Grant Nos. DE-FG02-96ER40989, DE-SC0026091, and DE-SC0023532, by the National Science Foundation under Grant No. PHY-2012522, and in part by the National Nuclear Security Administration through the Center for Excellence in Nuclear Training and University Based Research (CENTAUR) under grant number DE-NA-0004150. This work benefited from high performance computational resources provided by LSU (www.hpc.lsu.edu), the National Energy Research Scientific Computing Center (NERSC), a U.S. Department of Energy Office of Science User Facility at Lawrence Berkeley National Laboratory operated under Contract No. DE-AC02-05CH11231, as well as the Frontera computing project at the Texas Advanced Computing Center, made possible by the National Science Foundation award OAC-1818253. The authors thank A. Laird, R. Longland, and G. Sargsyan for helpful discussions.

\end{acknowledgments}

\bibliography{DecayChain}

\appendix

\end{document}